\documentclass[lettersize,journal]{IEEEtran}
\usepackage{cite}
\usepackage{amsmath,amssymb,amsfonts}
\usepackage{tikz}
\usepackage{color, soul}
\usepackage{pgf-pie}
\usepackage{algorithmic}
\usepackage{graphicx}
\usepackage{textcomp}
\usepackage{hyperref}
\usepackage{caption}
\usepackage{booktabs,tabularx}
\usepackage{enumitem}

\begin{document}

\title{Rumor Stance Classification in Online Social Networks: The State-of-the-Art, Prospects, and Future Challenges}

\author{Sarina Jamialahmadi,
		Iman Sahebi,
		Mohammad M. Sabermahani,
		Seyed P. Shariatpanahi,
        Aresh Dadlani,~\IEEEmembership{Senior Member,~IEEE},
        Behrouz Maham,~\IEEEmembership{Senior Member,~IEEE}
\thanks{This research was partly supported by the Faculty Development Competitive Research Grant\,\,(No. 240919FD3918), Nazarbayev University, under the project entitled ``Data-Driven Framework Development for Projection and Containment of Spreading Processes in Complex Social Networks''.}
\thanks{S.\,Jami,\,I.\,Sahebi,\,and\,S.\,P.\,Shariatpanahi\,\,are\,\,with the School of Electrical and\,Computer\,Engineering,\,College\,of\,Engineering,\,University\,of\,Tehran,\,Iran.}%
\thanks{M. M. Sabermahani is with the School of Computer Engineering, College of Engineering, Amirkabir University of Technology, Iran.}%
\thanks{A. Dadlani is with\,\,the\,\,Faculty\,\,of\,\,Arts,\,\,University\,\,of\,\,Alberta, Edmonton, AB\,\,T6G\,\,2E7, Canada, and also with the Department of Electrical and Computer Engineering, Nazarbayev University, 010000 Nur-Sultan,~Kazakhstan.}
\thanks{B. Maham is with the Department of Electrical and Computer Engineering, Nazarbayev University, 010000 Nur-Sultan, Kazakhstan.}}%

\maketitle

\begin{abstract}
The emergence of the Internet as a ubiquitous technology has facilitated the rapid evolution of social media as the leading virtual platform for communication, content sharing, and information dissemination. In spite of revolutionizing the way news is delivered to people, this technology has also brought along with itself inevitable demerits. One such drawback is the spread of rumors expedited by social media platforms, which may provoke doubt and fear. Therefore, it is essential to debunk rumors before their widespread use. Over the years, many studies have been conducted to develop effective rumor verification systems. One aspect of such studies focuses on \textit{rumor stance classification}, which involves the task of utilizing user viewpoints regarding a rumorous post to better predict the veracity of a rumor. Relying on user stances in rumor verification has gained significant importance, for it has resulted in significant improvements in the model performance. In this paper, we conduct a comprehensive literature review of rumor stance classification in complex online social networks (OSNs). In particular, we present a thorough description of these approaches and compare their performances. Moreover, we introduce multiple datasets available for this purpose and highlight their limitations. Finally, challenges and future directions are discussed to stimulate further relevant research efforts.
\end{abstract}


%



\section{Introduction}
\label{sec:introduction}
\fontdimen2\font=0.48ex
\IEEEPARstart{S}{ocial} media have undoubtedly matured as the foremost technology for browsing information, sharing opinions, and disseminating news. The widespread use of social media has diminished the limitations of cultural and geographical access to information. As a result, journalists leverage social media channels to find emerging news as soon as they happen~\cite{Zubiaga2013CuratingAC}. Additionally, online social networks (OSNs) have drastically changed how audiences reflect on the construction and distribution of media messages. According to~\cite{Hermida2012SHARELR}, social media has belittled the role of news authorities such that people are usually more likely to receive news from their friends and acquaintances rather than news organizations or journalists. Therefore, social media platforms have provided opportunities to share and interpret news liberally. Consequently, this affair has led to the rise of unverified news, which may confuse people and raise doubts. Although it has promoted lives, the shortcomings of online social platforms cannot be overlooked. Unverified information, typically called \textit{rumors} on OSNs, may later be determined as false. Pervasive false rumors, while their veracity is not yet known, may adversely affect individuals or communities~\cite{Zubiaga2014TweetBV}. Such form of information tends to mislead people or even sometimes provoke dangerous activities. For instance, the Pizzagate conspiracy, that went viral on social media in 2016, speculated the establishment of a pedophilia ring in a pizzeria located in Washington, D.C., which triggered an armed man to go to that restaurant. Not only did he not witness any crime at the scene, but was also arrested~\cite{wendling_2016}. Another example of the adverse impact of incorrect information is the rapid fall of the Dow Jones Industrial Average in 2013, immediately after the hacked Associated Press Twitter account posted a bogus tweet about an attack on the White House~\cite{murphy_2013}. Evidently, the need to detect rumors as soon as they are posted online greatly increases the chances of averting any potential harm.

Owing to the ever-increasing amount of content online, the manual verification of rumors is a time-consuming and labor-intensive process. This problem has prompted many researchers to develop automatic systems for detecting and verifying rumors on social media, prior to their extensive diffusion. Studying the nature of rumors has long been the focus of researchers from different disciplines, ranging from psychological studies to computational analysis~\cite{Zubiaga2018DetectionAR}. These studies have revealed interesting facts about rumor diffusion and users' contributions to it. Therefore, a framework for rumor verification is essential for alerting users to unreliable information and encouraging them to exercise greater caution by debunking and stopping the spread of false information. In what follows, we define misinformation and its various facets in order to provide a better perspective on the attributes of rumors, followed by our research focus, which is rumor detection as a component of this subject.

\subsection{Misinformation in OSNs}
\fontdimen2\font=0.48ex
Misinformation in social media has a wide range of definitions, as well as various detection techniques, datasets, and challenges. The terms misinformation, disinformation, and malinformation are frequently used interchangeably despite the subtle differences between them. Misinformation refers to false or misleading information with no intention to harm or deceive. Disinformation is deliberately fabricated or biased information with the intent to deceive and cause harm. Malinformation is genuine private information that is intentionally spread for personal or corporate interests. \figurename{~\ref{fig1a}} {differentiates the three terminologies based on information validity and user intention. Before delving into rumor detection, we first identify the various forms of misinformation and disinformation shown in \figurename{~\ref{fig1a}}.
\begin{figure}[!t]
    \centering
    \includegraphics[width=1\columnwidth]{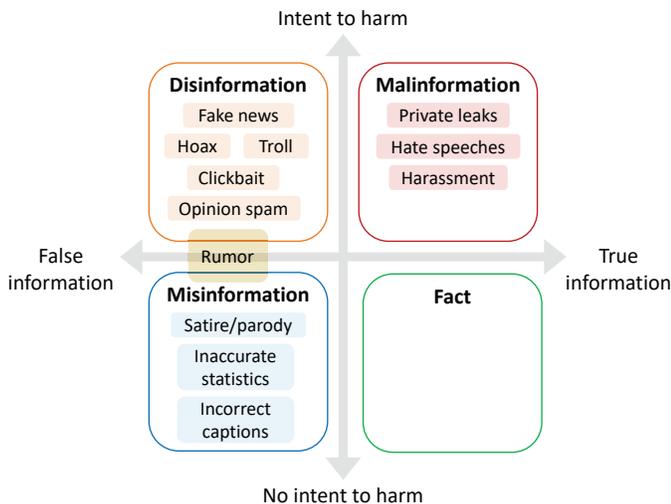}
    \caption{Information disorder spectrum.}
    \label{fig1a}
    \vspace{-0.8em}
\end{figure}

Individuals or groups spreading misinformation on social media may have several intentions. Despite the fact that determining the purpose of a misinformation diffuser is non-trivial, misinformation itself can be categorized into two types based on the objective of the spreader \cite{LWu_2019}. False information that is intentionally diffused by some people and groups who have the motive to deceive and mislead individuals or societies is referred to as \textit{intentionally-spread misinformation}. Contrarily, disinformation that has been inadvertently transmitted over the Internet due to unawareness or lack of confidence of regular users in the source is referred to as \textit{unintentionally-spread misinformation}. The goal of these users is to inform others and not to mislead them.

\textbf{Rumors} are unverified information that spread across social media platforms, which may later be determined as true or false or remain unverified. Rumors are usually short posts on social media that spark debate among users in order to establish their veracity. Rumors can be broadly classified into \textit{long-standing rumors} and \textit{emerging news}~\cite{Zubiaga2018DetectionAR}. Long-standing rumors refer to stories about celebrities or public figures that circulate for a long time on social media without being accurately verified. Emerging news is defined as news that propagate during a crisis or important events that have not previously been observed. In this survey, we examine the strategies developed to date for rumor detection in social media. \textbf{Satire/parody} refer to articles that are primarily humorous and ironic, with no malicious purpose yet with the capacity to fool. Another form of misinformation is \textbf{erroneous messages} in statistical reports and captions of images and tables.

Unlike rumors, \textbf{fake news} are news pieces published with the intent of deceiving people. They emerge in the context of serious fabrications of content (like deep fake)~\cite{Bondielli201938}. Serious fabrications refer to spiteful news articles that tend to deceive people and often propagate rapidly on social media. Large-scale \textbf{hoaxes} are incorrect news stories that target public figures or ideas on a wider scale than typical news. \textbf{Opinion spams} refer to fake or biased reviews or comments that aim to promote or impede a product or services. There are two types of opinion spam, i.e., \textit{hyper spam} and \textit{defaming spam}. Hyper spam refers to positive reviews that are fabricated to elevate a product, while,  defaming spam aims to negatively affect the customers of a product \cite{su_qimotive}. \textbf{Clickbait} refers to messages designed to entice readers to click on a link by evoking their emotions and interests. This kind of writing is typically employed for website advertising, however studies have shown that clickbait is a key source of falsified information transmission on social media~\cite{Elyashar_2022}. \textbf{Troll} refers to a person who wants to manoeuvre people to change their opinions in a discussion or to make them angry by acting in a violent manner. Trolls are known to have an annoying behavior and try to disrupt communities. In extreme cases, trolls can potentially harm people by tricking them into communicating their feelings and private information, which can be upsetting to the victim~\cite{fi12020031}.

Although some studies contend that rumors and false information are quite similar, it is essential to differentiate between each one and show how it varies from the other. According to \cite{Bondielli201938}: ``\textit{they are strongly connected, and may be considered as different aspects of the same general problem}''. As discussed in \cite{Bondielli201938}, fake news and rumors are different in their definitions, available datasets, methods of collecting data, and extracted features. While fake news refers to news stories that are frequently seen on harmful news websites, rumors are thought of as social media posts. Because of this, there are differences between the public datasets and the techniques used to gather and annotate these two types of false information, which also causes some features to be platform-specific. Despite being different in their definitions and datasets, the techniques to extract features and design models can be similar for rumors and fake news detection systems. In this paper, we merely focus on rumor stance classification and highlighted the results of the reported models on rumor datasets. Although some reported works have trained their models on both rumors and fake news datasets, we believe that most of these methods and Natural Language Processing (NLP) analysis can be also applied to a fake news detection. However, in terms of usefulness and effectiveness, we cannot declare with certainty that a method suggested for a rumor detection system produces the same results for fake news detection.

We now introduce some approaches that have been widely adopted in misinformation detection on OSNs. \textbf{Stance detection} is the practice of determining users' standpoint towards a target or a claim. Since stance detection does not judge the truthfulness of a claim but rather gives information about the author's point of view to aid in the verification of the assertion, it differs from rumor detection and fake news identification. Social media users participate to determining the truthfulness of news by debating the evidence that supports or contradicts the report \cite{10.1145/3369026}. Due to the ever-increasing amount of content being shared online and the necessity to look into the problems that are emerging as a result, stance detection has become a growing study area. In this work, we investigate how stance detection can be used to improve rumor verification system performance. The term \textit{rumor stance classification} is typically used to describe stance detection in rumor detection systems, i.e., users' judgments about a rumorous post are used to validate rumors, which might help a rumor verification system.

\textbf{Sentiment analysis} is the practice of mining emotions from texts. In other words, whereas stance detection determines a user's perception of an entity, sentiment analysis may identify the positive and negative emotions present in the text. Sentiment analysis can be used to identify false news and other sorts of misinformation since their authors often employ heightened or opposing emotions to draw readers' attention \cite{su_qimotive}. \textbf{Fact checking} is an approach used in misinformation detection which compares the content of the unverified news with the actual facts derived from the Web (such as Wikipedia). This technique comprises of two stages: (i) construction of a knowledge graph by extracting facts from open Web and (ii) assessment of the content of the unverified information using the constructed knowledge \cite{10.1145/3395046}.
\begin{figure*}[!t]
    \centering
    \includegraphics[width=0.9\textwidth]{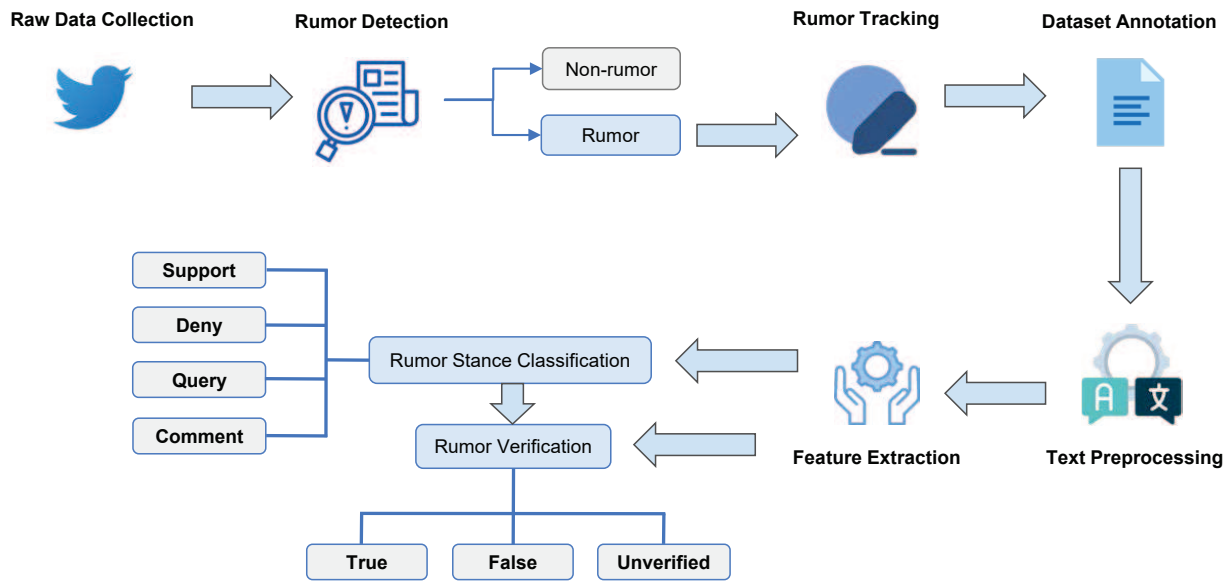}
    \caption{Schema of the development stages in a general rumor detection system.}
    \label{fig2}
\end{figure*}

\subsection{Definitions of Rumors and Stances}
\fontdimen2\font=0.52ex
According to \cite{Zubiaga2018DetectionAR}, rumor is defined as ``\textit{an item of circulating information whose veracity status is yet to be verified at the time of posting}''. That is to say, rumors are pieces of information diffusing through social media platforms that do not hold definite veracity. As evidence pertaining to the rumorous claim is gathered over time, the veracity of the rumor can be determined. In some cases however, the existing evidence is not sufficient to ascertain the rumor veracity and thus, remains unverified \cite{Zubiaga2016AnalysingHP}. As a result, rumorous posts are typically labeled as either \textit{true}, \textit{false}, or \textit{unverified} \cite{Zubiaga2016AnalysingHP}.

Moreover, social media users tend to discuss the veracity of a rumor and express their standpoint on the rumorous claim. For instance, they might question the credibility of the rumor and ask for a reliable source; they may favor or oppose the rumor and mention a source to manifest their statement; or they can simply remark on the issue without showing approval or otherwise \cite{10.1145/3132847.3133116}. These explanations comprise the categorization of users' stances, i.e., \textit{query}, \textit{support}, \textit{deny}, and \textit{comment} \cite{Zubiaga2018DetectionAR}. From a linguistic point of view, stance is defined as ``\textit{a public act by a social actor, achieved dialogically through overt communicative means of simultaneously evaluating objects, positioning subjects (self and others), and aligning with other subjects, with respect to any salient dimension of the socio-cultural field}'' \cite{englebretson2007stancetaking}.

The three main definitions for stance detection given in \cite{10.1145/3369026} are: (i) \textit{generic stance detection} that determines the stance of a text towards a specific target, (ii) \textit{rumor stance classification} in which the opinion of the author of the post towards the rumorous story is captured, and (iii) \textit{fake news stance detection} where there exist headline and body of the same or different news articles and the stance of the headline is determined towards the body. rumor stance classification is undoubtedly, a developing and favored research topic \cite{10.1145/3369026}. A collection of user stances in discussions around a rumorous post can yield essential information about the rumor \cite{Derczynski2014PHEMEC}. 

In \cite{Zubiaga2018DetectionAR}, a rumor detection framework is described as a bodywork consisting of four components, namely \textit{rumor detection}, \textit{rumor tracking}, \textit{stance classification}, and \textit{veracity classification}. The first two components are requisite modules that collect rumorous posts and related conversations. Simply put, the rumor detection component identifies whether a post is rumor or non-rumor so as to mark rumorous claims. The rumor tracking component is then employed to collect conversations related to the rumorous claim. Upon data collection, the claim and the conversation around it are used in building two classification systems. One is stance classification (the third component) which determines the orientation of related posts towards the source rumor; This task is usually referred to as rumor stance classification \cite{Zubiaga2018DiscourseawareRS}. The other is veracity classification module (the fourth component) which assesses the veracity of the rumor. In many studies, the output of stance classification serves as input for veracity classification systems. As shown in \figurename{~\ref{fig2}}, raw data is first collected from social media platforms and a rumor detection method is applied to determine if the source post is a rumor or not. Then, in the rumor tracking phase, conversations around the rumorous claims are gathered and the dataset is annotated. Subsequently, NLP methods are employed to pre-process the dataset and useful features are extracted to prepare inputs for either a rumor stance classifier or a rumor verification classifier. As discussed before, the output of the rumor stance classification can be used as features in a rumor verification system. Our main objective is to survey the existing works and progress on \textit{rumor stance classification} reported to date.

\subsection{Background}
\fontdimen2\font=0.52ex
The study of rumors spans over a broad range of disciplines, from the mathematical analogy of rumor propagation to infection spread \cite{watson1987size}, assessing the psychological aspect of rumor in a community and the reasons for its origination \cite{+2014+35+47}, and applying computational methods to detect rumors over the Internet. A myriad of research works centered on online misinformation diffusion have been reported in recent years. These studies approach the problem by applying various methods such as rich feature set selection, usage of supervised and semi-supervised algorithms, temporal characteristics adoption, network propagation considerations and the like. Also, it has been shown that some specific patterns in tree-structured conversations can help demonstrate if a rumor is true or false \cite{semeval2017}. The first study of automatic rumor detection initiated by Qazvinian \textit{et al.} \cite{Qazvinian2011RumorHI} involved the collection of a dataset of long-standing rumors (rumors that circulate in social media for a long time without being accurately verified \cite{Zubiaga2018DetectionAR}), followed by the extraction of content-based features such as lexical and part of speech patterns, network-based features such as retweets, and Twitter-specific memes such as hashtag and URLs which were fed to a L1-regularized log-linear model. Furthermore, they exploited stances of users and defined this task as \textit{belief classification}, wherein users' attitudes towards the rumorous claim were classified in three categories, namely \textit{confirm}, \textit{deny}, and \textit{doubtful}. The class `comment' was excluded, and the class `doubtful' was combined with the class `question' in the new classification of rumor stances. Their work was further extended in \cite{Cai2014RumorsDI} to study long-standing rumors on social media.

This approach of collecting data, however, failed to capture rumors not known \textit{a priori} thus, limiting the understanding of the nature of rumors on social media. In 2016, Li \textit{et al.} \cite{Zeng_Starbird_Spiro_2021} collected a dataset of crisis-related messages from Twitter and justified the verification of emerging misinformation in reducing the misuse of social media for crisis communication, unlike previous work that focused only on rumors about prominent profiles. They proposed rumor stance classification models using Random Forest, Logistic Regression, and Naive Bayes with two classes, i.e., \textit{affirm} and \textit{deny}. Features such as punctuation, Twitter-element, lexical, and tweet sentiment were extracted manually and all models were trained on several different training data samples, i.e., event-based datasets and pooled datasets (combining all events). At the same time, Zubiaga \textit{et al.} \cite{Zubiaga2016AnalysingHP} embarked on a methodology that could identify rumors not known previously. They collected a diverse set of newsworthy rumors and performed a quantitative analysis on rumors propagation and users' reaction. They illustrated their observations on diffusion of rumors in social media network and users' tendency to support or deny rumors. For instance, tweets that support a rumor that is still unverified have a much higher number of retweets compared to verified news, but after a rumor is validated, the number of retweets greatly decrease regardless of whether it is a true or false. In addition, users are disposed to support every unverified rumor. After a resolving tweet is published, discussions around that specific rumor increase and if the rumor turned out to be false, the denial tweets exceed the supporting tweets, while in other cases the denial messages are usually lower than the supporting messages in number.

\subsection{Motivation}
\fontdimen2\font=0.52ex
This survey paper is motivated by the unresolved problem of the increasing diffusion of rumors in social media and the potential harm it poses upon individuals and societies. Furthermore, the utilization of users' stances have shown promising results in the verification of rumors, and a large variety of approaches have been adopted towards exploiting users' stances. In spite of the existing surveys on rumor detection systems \cite{Hardalov2021ASO, Pathak2020AnalysisOT, Varshney2021ARO}, a specifically comprehensive review on the recent methods, features, datasets, and challenges for \textit{rumor stance classification} has not yet been reported. To fill this gap, this paper concerns the state-of-the-art in rumor verification systems by exploiting wisdom of the crowd. Thereby, we conduct a comprehensive survey over these studies to provide insights on the development of a rumor stance classification system for future researchers and practitioners. Our paper mainly focuses on the proposed methods and findings of a rumor stance classification system. We also discuss the various datasets and different categories of feature sets that are widely considered for rumor stance classification.

\subsection{Paper Structure}
\fontdimen2\font=0.52ex
The remainder of this paper is organized as follows. Section~\ref{sec:related work} introduces the related works on rumor stance classification. The approaches are then comprehensively discussed in Section~\ref{sec:rumorMethod}, which describes various traditional and modern learning algorithms for rumor stance classification, and Section~\ref{sec:timeconsider}, which introduces approaches that adopt time of detection as a major factor of their models' performances. In these two sections, each method is explained in terms of the adopted features and the merits of the model. An assessment of feature sets and their usefulness is discussed in Section~\ref{sec:featuresets}. In Section~\ref{sec:datasets}, several datasets that are either publicly used or specific to a research work are presented along with explanation on their characteristics. This is followed by the evaluation metrics and their likelihood of revealing model's performance on all classes in Section~\ref{sec:evaluation}. Section~\ref{sec:findings} unfolds findings on the stances of the microblog users. Finally, the potential research directions are highlighted in Section~\ref{sec:future}, followed by the conclusion in Section~\ref{sec:conclusion}.

\section{Related Work}
\label{sec:related work}
\fontdimen2\font=0.52ex
The expansion of research on rumor verification began with the introduction of datasets for rumor stance classification and rumor veracity prediction at the Semantic Evaluation (SemEval) workshops held in 2017 and 2019 \cite{semeval2017} \cite{semeval2019}. Global participation in these competitions encouraged more researchers to develop out-of-competition models and use the models from participants of SemEval as baselines to compare the performance of their methods. In this section, we look at rumor stance classification approaches developed at the SemEval competitions in years 2017 and 2019.

In SemEval 2017, Task 8 was dedicated to rumorEval which aimed to develop methods to verify rumors on social media. This task included two sub-tasks: (A) for rumor stance classification, and (B) for rumor verification. There were eight participants for sub-task (A) with the winning system team `Turing'. Kochkina \textit{et al.} \cite{Kochkina2017TuringAS} (Turing) utilize word2vec, tweet lexicon, punctuation, content length, and relation to other tweets as main features and apply a Branch-LSTM model to extract information from sequences of tweets. Their model misclassifies all instances of \textit{deny} class which they mention by the small number of data points for this class. Bahuleyan \textit{et al.} \cite{Bahuleyan2017UWaterlooAS} adopt manual feature engineering to make use of cue word features like skepticism, knowledge, and denial, as well as message-specific features. They apply a gradient boosting classifier (XGBoost), train it with various feature sets, and find that using all features produces the best results. Wang \textit{et al.} \cite{Wang2017ECNUAS} employ ensemble method with majority voting. To address the imbalance dataset problem, they offer a two-step classification methodology in which `comment' instances are categorized first, followed by classification of the three remaining classes, i.e., \textit{support}, \textit{deny}, and \textit{query}. Similarly, Lozano \textit{et al.} \cite{Lozano2017MamaEA} employ the ensemble method in conjunction with a voting strategy that assigns different amounts of importance to classifiers. They use convolutional neural networks (CNN) and a set of hand-written rules that value precision above recall. Chen \textit{et al.} \cite{Chen2017IKMAS}, too, implement a CNN with the Softmax classifier that gives a probability to each of the classes. Enayet \textit{et al.} \cite{Enayet2017NileTMRGAS} and Singh \textit{et al.} \cite{singh2017} train support vector machine (SVM) classifiers using features like hashtags and user-based features. Srivastava \textit{et al.} \cite{Srivastava2017DFKIDKTAS} train three models, including maximum entropy, naive Bayes, and winnow classification, before using a majority vote classifier to get the best results. They extract features specific to each category of stances when it comes to features.

In the SemEval workshop held in 2019, a new dataset from Reddit platform was combined with the dataset of SemEval 2017 and used for rumor evaluation in Task 7. Task (A) had 22 participants, with three systems (BLCU-NLP \cite{yang-etal-2019-blcu}, BUT-FIT \cite{Fajcik2019BUTFITAS}, eventAI \cite{eventAI2019}) earning macro F1-scores of more than $50\%$. Yang \textit{et al.} \cite{yang-etal-2019-blcu} use other datasets to expand the dataset on three minority classes: \textit{support}, \textit{deny}, and \textit{query}. They extract Twitter and Reddit-specific features, as well as concatenating conversation context and target tweet for the input to include information from earlier tweets. They use OpenAI GPT from prior research and find that it works best with an input that includes source, parent, target, and other tweets, as well as word-level feature representation. Fajcik \textit{et al.} \cite{Fajcik2019BUTFITAS} use a pre-trained BERT model and train $100$ models with varied learning rates. They run the experiment with several settings and include any improvements brought about by various parameters into the model. Li \textit{et al.} \cite{eventAI2019} point out that the dataset contains contradictory labels based on inter-annotator agreement rates, and in order to resolve the issue, they discover similar postings with various labels and apply the same label to all of them. They deploy an ensemble classifier that includes a neural network with two fully connected layers, SVM, random forest, and logistic regression, as well as a collection of features such as relation to source message and link type (i.e., picture, video, article).

The variety of models and feature sets used in these competitions can be used to expand the research in this area and develop a rumor detection system that can be applied to a real-world platform with high speed and accuracy. In what follows, we categorize efforts concerning methods of rumor stance classification and practical rumor detection systems into the following two sections:
\begin{itemize}
    \item \textit{Rumor Stance Classification} which includes two approaches for labeling stances, i.e., \textit{Traditional Machine Learning} and \textit{Deep Learning}. The former refers to methods that use traditional learning algorithms, and develop and produce rich feature sets in order to improve the performance of the models. The latter employs methods such as CNN and long short-term memory (LSTM) in single-task learning, i.e., only training rumor stance classification task, and multi-task learning, i.e., utilizing various neural networks to learn rumors and stances jointly.
    \item \textit{Rumor Detection} which includes two frameworks, i.e., \textit{Early Detection} and \textit{Real-Time Systems}. Early detection aims to detect rumors as soon as they are published which struggles with few data available in early hours of rumor propagation. Real-time systems are built with the purpose of collecting and processing the data in real-time and giving users a dynamic platform where they can search rumors and obtain insights about their veracity.
\end{itemize}

\section{Rumor Stance Classification}
\label{sec:rumorMethod}
Various approaches have been studied to develop an effective and coherent rumor detection framework ranging from traditional algorithms to neural networks. In this section, we study these methods in terms of the adopted features and the merits of the model.

\subsection{Traditional Machine Learning}
\label{sec:conventional}
Numerous research efforts have demonstrated that by incorporating more useful features into simple learning models, great performance can be achieved. Pamungkas \textit{et al.} extract features indicating Twitter language, conversation-based features, and affective-based features which reveal emotional aspect of tweets content \cite{Pamungkas2018StanceCF}. They apply these features to an SVM classifier and show that conversation features have a significant impact on model performance without the requirement for complex learning procedures. Moreover, Aker \textit{et al.} \cite{Aker2017SimpleOS} use the $J48$ decision tree model to a number of automatically identifiable problem-specific features, such as computed scores for uncertainty, certainty, and so forth. The authors demonstrate that each of these characteristics contributes to the performance improvement of the model. In another work \cite{Aker2017StanceCI}, the authors focus on the stance classification of rumors related to mental health disorders. They collect a small sample of health-related tweets and supplement it with additional data from the PHEME dataset, but these are not health-related tweets. They divide training and testing data in two ways: (i) data in the testing set has never been observed before (is not included in the training set), and (ii) a part of data from the testing set ($10\%$ to $60\%$) is included in the training data. SVM, J48, random forest, naive Bayes, and instance-based classifier are the learning models used. They first use the two techniques on non-health data, then take the health data as the testing set and train all the classifiers on non-health data as a second approach. In both scenarios, J48 performs best, however when the instances in the testing data are not observed by the model, the accuracy and macro-average F1 score decline, implying that different types of rumors have different features that cannot be caught by a model that has not seen the data before.

Lukasik \textit{et al.} \cite{Lukasik2019GaussianPF, lukasik-etal-2015-classifying} employ Gaussian processes classifiers on two datasets, England riots and PHEME, and use two approaches, Leave-One-Out and Leave-Part-Out (the same as Aker, but with a maximum of $7\%$ of target rumor included in the training data). On the England riots dataset, they discover that Gaussian Processes detect the distribution of distinct stances more successfully than other baseline models. The same is true for the PHEME dataset, but the result is lower. This could be due to the presence of different events in the PHEME dataset, so the classifier encounters features that were not seen in the training data, and also because the PHEME dataset consists of tweets that are replies to other tweets and may be shorter, causing the classifier to struggle to understand the stance of the tweet. In another paper, Lukasik \textit{et al.} train the Hawkes process classifier on the PHEME dataset using temporal information, demonstrating the utility of temporal variables in the classification of stances \cite{Lukasik2016HawkesPF}. Dungs \textit{et al.} use stances within a Hidden Markov Model (HMM) to check rumors in order to demonstrate the efficiency of users' stances for determining the truth of rumors \cite{romourstancealonepredictveracity}. In another case, the authors use multi-spaced HMM to account for time as well. They demonstrate that incorporating collective stances as characteristics added to other extracted features in the model can assist improve the performance of rumor detection models.

Giasemidis \textit{et al.} \cite{semisupervisedGias} claim that semi-supervised algorithms outperform supervised methods on large-scale data. Due to limitations in supervised algorithms, it is an ineffective tool for capturing message attitudes toward various rumors. Furthermore, hybrid supervised algorithms are time-consuming and expensive because the model must be retrained for each new rumor, whereas unsupervised models are insufficient because we cannot acquire specifics of each cluster output by the algorithm. Semi-supervised algorithms, as noted in the study, require less manual labor and are faster and rumor-specific, i.e., they classify texts based on the content of the individual rumor. As additional annotated data becomes available, the authors use label spreading and label propagation algorithms, resulting in a faster and more accurate system. They assess the model using four distinct feature set settings in order to select the best performing system. The end-user in this system must name a few of the first messages of a rumor to feed to the algorithm, which is a hurdle when the user requires the result instantly, which is addressed in the study by presenting a supervised solution.

\subsection{Deep Learning}
\label{sec:neuralnetwork}
Neural networks constitute a considerable portion of automatic learning methods. These methods are vastly used in the development of rumor stance classification systems which have yielded remarkable results. In this section, we study two approaches in rumor stance classification which utilize neural networks, i.e., single-task learning and multi-task learning.

\subsubsection{Single-Task Learning}
\fontdimen2\font=0.52ex
Deep learning has grown in popularity as a feature extraction method since it involves less manual work than traditional machine learning methods. Zubiaga \textit{et al.} \cite{Zubiaga2016StanceCI} develop sequential classifiers with only local features in order to capture the effect of conversational threads on model performance. They compare models with sequential and complete conversational tree structures using tree and linear-chain conditional random fields (CRFs). In a subsequent study \cite{Zubiaga2018DiscourseawareRS}, driven by the effects of tweets on one another, they expand on the previous study by integrating more sequential classifiers and leveraging dialogical structure in tweets to analyze the effect of Twitter conversational thread on both sequential and non-sequential classifiers. They find that contextual characteristics significantly improve the performance of non-sequential classifiers, but have little impact on sequential classifiers since these classifiers can learn the context in conversational threads inherently. As a result, while adding contextual data to sequential classifiers may increase redundancy, sequential classifiers outperform non-sequential classifiers without contextual features.

Poddar \textit{et al.} \cite{Poddar2018PredictingSI} propose a real-time system model, wherein CNNs are used to extract textural information from tweets, while recurrent neural networks with a tweet-level attention mechanism are used to account for the sequence of tweets in a conversation tree. They use timestamped information to predict stances and average the output of each stance class for the entire tree to detect the veracity of rumors. Veyseh \textit{et al.} \cite{10.1145/3132847.3133116} extract contextual features with temporal attention and relation traits. They gain aid from previous and next tweets of the current tweet for temporal features, then employ a window of six tweets - three for previous and three for following tweets - and apply a CNN to extract features from this configuration. Ma \textit{et al.} \cite{Ma2018RumorDO} use a tree-structure recursive neural network with structural and content information integration to reinforce or weaken a node's stance via recursion in the propagation tree. They offer two model settings: (1) bottom-up tree (the direction of link points from the response node to its responded node) and (2) top-down tree (the direction of link points from the responded node to the response node). For rumorous postings, the top-down model beats other models; however, for non-rumorous posts, these two models fail to perform well, and other feature engineered models outperform them, indicating that more information, rather than content and structure, is required, such as user information.

In \cite{Bai2020ASA}, Bai \textit{et al.} propose a stochastic attention mechanism using CNN for feature extraction, underscoring that the traditional attention mechanism could not identify crucial information in users' messages due to differences in reading and writing habits. Their model can discover classification-related keywords and exclude non-classification-related terms across the whole connection layer. Kumar \textit{et al.} \cite{kumar2021rumor} use the output of two neural networks to forecast user stances: Capsule neural network, which accepts texts as fixed-length vectors, and Multi-Layer Perceptron neural network, which takes features extracted from texts. They indicate that the \textit{query} occurrences are largely properly predicted, highlighting the value of the Multi-Layer Perceptron network. They also compare the performance of Capsule network with CNN by substituting the CNN with the Capsule network in the same setting, and find that when Capsule network is used, the overall average F1 score increases by $4\%$.

The authors of \cite{Scarton2021CrosslingualRS} have suggested a cross-lingual rumor stance classification on four languages, including English, German, Danish, and Russian, in order to broaden rumor stance classification into different languages. They use a multilingual BERT model (which addresses data imbalance by adjusting thresholds) and Machine Translation in four different settings: (1) \textit{Monolingual model}, which trains MBERT for each dataset separately. They also machine translate Danish and Russian, and train MBERT model on them as well. (2) \textit{Zero-Shot}, in which they use an MBERT trained on an English dataset for other languages. This is also applied to machine translations of German, Danish, and Russian. (3) \textit{Few-Shot} which takes MBERT trained on English dataset and fine-tunes it with target languages. (4) \textit{Full multilingual model} that is jointly trained on all languages. The authors discover in this experiment that few-shot models outperform multilingual models on all classes for Danish and Russian, possibly because the model particularizes the target language in the few-shot condition. Khoo \textit{et al.} \cite{khoo2020interpretable} emphasize that users in a conversation thread may not be responding to another user, but rather to the thread as a whole. As a result, they create a post-level attention model with two variants: a post-level attention mechanism that takes tree structural information into account, and a structure aware attention mechanism with a hierarchical attention model at token-level and then at post-level to learn a more complex sentence representation. They also consider time delay in the importance of tweet replies, explaining that skepticism expressed early in the rumor's propagation may not be of much value because the claim has not yet been verified, whereas skepticism expressed later in the rumor's spread may be of high importance due to the availability of more information. They employ the PHEME and Twitter $15$ and $16$ datasets \cite{inproceedings-twitter15&16}, and they observe variable outcomes across datasets, with a performance decrease on the PHEME dataset. One issue is the way training and test data are split; for example, PHEME splits data at the event level, whereas Twitter $15$ and $16$ data points are split arbitrarily.

\subsubsection{Multi-Task Learning}
\fontdimen2\font=0.52ex
Because postures might provide useful information for determining the validity of a rumor, multi-task learning can be a viable choice for veracity prediction \cite{Wei2019ModelingCS}. Multi-task learning is advantageous in various respects, including getting more data for the training set and reducing the likelihood of over-fitting on a single dataset \cite{kochkina2018all}.

Kochkina \textit{et al.} \cite{kochkina2018all} employ a BranchLSTM to accomplish joint multi-task learning in three ways: stance classification and rumor verification, rumor detection and rumor verification, and stance and rumor detection with rumor verification. They demonstrate that the multi-task learning model beats competing models and baselines on all three tasks. The models, however, perform differently for each class label. For example, single task learning models are better at predicting false rumors, whereas multi-task learning models are better at predicting unconfirmed rumors. Ma \textit{et al.} \cite{10.1145/3184558.3188729} propose two multi-task learning models. The first model has a Uniform Shared-Layer Architecture which includes a single shared hidden layer, and the second model has an Enhanced Shared-Layer Architecture comprising additional task-specific hidden layers. They point out that most studies so far have not considered the stance classification task as a sequence problem. This problem is addressed by taking into account the contextual information of articles (from fake news dataset) or posts (from rumor dataset) associated to a particular headline or a claim. In addition to their two multi-task learning models, they also train a single-task learning model by removing the rumor detection task from the Uniform Shared-Layer Architecture which could superbly show the significant improvements made by multi-task learning in the evaluation results. To indicate the noteworthy contribution of multi-task learning in misinformation detection, they conduct an experiment to find the important patterns from the hidden layers. The results show that the task-specific layer can capture patterns that is also captured by the single-task learning, the shared-layer can capture patterns that may enhance the performance when is applied with the task-specific patterns, and that the outputs of the task-specific and the shared-task layers have little in common which can be interpreted as the two layers can work together and enhance each other.

Poddar \textit{et al.} \cite{Poddar2018PredictingSI} use three methods to use stance prediction results in the rumor verification task: pipeline model, which uses the output of the stance prediction model as input for veracity detection; joint model, which applies a multi-objective loss function on a single model; and transfer learning, which uses the stance prediction model weights for the entire model. The results reveal that transfer learning produces higher accuracy, which could be attributed to two factors, as stated in the paper: (1) the dependency between stance and veracity prediction tasks, and (2) an imbalance in dataset sizes. Khandelwal \textit{et al.} \cite{finetunelongformer} employ Longformer Transformer with a sliding window-based attention mechanism to collect user stances in order to learn the development of stance in a dialogue. The stance classification and rumor verification tasks are learned together, and the ensemble approach is used to obtain the maximum F1-score from a pool of models with alternative configurations. Islam \textit{et al.} \cite{Islam2019RumorSleuthJD} propose a multi-task learning model based on recurrent neural networks, with one shared layer and task-specific layers. They utilize a variational autoencoder to represent the user's features and point out that using the shared layer output for stances is not a proper way of feature representation; thus, this output is fed to an LSTM network linked to the stance classification task.

More recently, Zhang \textit{et al.} \cite{9376933} propose some solutions to improve joint learning stance classification and rumor verification, namely: (1) the data in both tasks may contain different modalities, such as image and text, and different modalities add to each other, so considering modalities other than text would improve performance; (2) the two tasks have different feature spaces due to differences in labeling, so applying generic feature-sharing multi-task learning would cause problems; (3) stance labels contain semantic information that can be used to boost performance by assigning weight to distinct stance labels. The authors propose a multi-task learning method based on hyper meta-networks that makes use of meta knowledge from both tasks. Their approach includes a multi-modal content layer, a shared layer for capturing meta knowledge from both tasks, and a task-specific output layer for assigning weights to distinct stances via an attention mechanism. Furthermore, Kumar \textit{et al.} \cite{kumarlstm} compare three approaches to LSTM employing multi-task learning, namely Branch LSTM, Tree LSTM, and a Binarized Constituency Tree LSTM, where a tree structure with two children for each node streamlines the matrix computations required for neural networks. Because of the difficulty in balancing stance classes with this alternative structure, this binarized constituency representation for tree structure did not perform well for rumor stance classification, and it appeared comparable for rumor verification task, which could be for a multitude of nodes created after binarization, making information propagation difficult.

In another relevant work, Yu \textit{et al.} \cite{Yu2020CoupledHT} describe how earlier approaches to multi-task learning did not take inter-task dependencies into account while using stance labels for the rumor verification task. They propose a Hierarchical Transformer that uses a BERT model to encode sentences. The authors highlight two existing problems with the BERT model in the task of rumor stance classification. The first is that we need sentence-level representation to understand conversations, whereas previous studies derive token-level representation with the BERT model. The second is that the BERT model has a limitation on the length of the sequence in the pre-training stage, and in our case, we have long conversation threads. To get over these limits, they cut the extensive sequences of the conversation threads into shorter ones. They then present a Coupled Transformer Module containing a stance-specific transformer and a cross-task transformer with a post-level attention layer to model the importance of each post, as well as the above-mentioned single-task model as low-level layers shared between rumor stance classification and rumor verification tasks. In a recent work, Yang \textit{et al.} \cite{Yang2022AWS} propose a weakly-supervised joint learning framework based on Multiple Instance Learning (MIL). The workflow of a MIL-based classifier entails categorizing the instances in a bag, such as sentences in a text, before merging the projected values from the instance-level to produce predictions at the bag level. The limitations of this classification approach are that rumor stance categorization and rumor verification are multi-class problems with independent labels, whereas the labels in the two classification tasks for an original MIL-based classifier are both binary and homogeneous. To address these challenges, they define multiple MIL-based binary problems in which they consider different veracity-stance class pairs as target classes for the binary classification problem, e.g., false-deny pair represents a positive class at the bag and instance levels and other pairs represent negative classes. Furthermore, they define two types of propagation trees for each claim, i.e., bottom-up and top-down. In the bottom-up model, stance of each post is predicted using information from all the children nodes and the veracity of the claim is determined using the aggregation of the stances by applying a bottom-up tree attention mechanism to obtain a better result. In the top-down model, which can better understand the typical pattern of information passing to each user and affect their stances, they design a top-down tree attention mechanism which selects the conclusive stances with their propagation path to determine the validity of the rumor.
\vspace{-0.2em}

\section{Rumor Detection}
\label{sec:timeconsider}
The methods we studied so far use datasets which consist of rumorous posts and their related tweets from any time frame after the rumor is published. Most of these methods do not consider the efficiency of their models in the early hours of the dissemination of a rumor. In this section, we study papers that design a rumor detection system which utilize users' stance and aim to detect rumors as soon as possible using few data available. In addition, we introduce real-time systems which collect data and extract features and users' stances in real-time such as websites which require them to be dynamic and fast to respond to a user's quest about rumors.

\subsection{Early Detection}
\fontdimen2\font=0.52ex
One key problem in rumor detection is detecting rumors as soon as possible before they cause harm to individuals or society. At the time of the initial rumor's publication, information to enable rumor verification is insufficient, thus further inquiry is required to help uncover rumors early. In this section, we study researches claiming their methods can detect rumors in the early hours of their propagation with a higher performance compared to other approaches that have not considered their methods to be efficient in early detection.

To validate rumors, the authors of \cite{Tian2020EarlyDO} depend solely on tweet content and early user reactions. They apply CNN and deep neural models based on  BERT to learn the stances of users' comments using transfer learning on the SemEval $2016$ stance detection dataset \cite{mohammad-etal-2016-semeval}, despite the fact that annotations for user stances do not exist. They next compare the performance of their system on Twitter $15$ and $16$ datasets to other models that have not employed transfer learning; their BERT-based model outperforms all other systems, particularly in the time close to the original rumor being posted. Ma \textit{et al.} \cite{inproceedings-twitter15&16} model rumor propagation trees and employ a kernel-based method to find similarities across these trees utilizing user information, context, and temporal features. They evaluate their models in the early hours of rumor dissemination, and while the propagation information is less in the early hours compared to later hours, their model outperforms alternative baselines that may use propagation information but cannot learn features from propagation structure well. According to their findings, rumors have a more complex dissemination structure than non-rumors in the early phases. A multi-spaced Hidden Markov model with stances and time as features may reach a macro F1-score of $61.8\%$ when examining the first $5$ tweets for the task of rumor verification, but this model achieves a macro F1-score of $80.4\%$ when using all tweets with an average of $18$ tweets \cite{romourstancealonepredictveracity}. The authors of \cite{Ma2018RumorDO} also accomplish early rumor detection using a tree-structure recursive neural network, demonstrating that comparable patterns of false rumors may be noticed in the early stages of a rumor being posted.
\begin{table*}[!t]
    \caption{Feature Description and Characterization}
    \label{tab:featureset}
    \setlength{\tabcolsep}{3pt}
    \centering
    \renewcommand{\arraystretch}{1.3}
    \begin{tabular}{|p{1.0\columnwidth}|p{1.0\columnwidth}|}
    \hline 
    \textbf{Textual Features} \cite{Zubiaga2018DiscourseawareRS} \cite{Aker2017SimpleOS} \cite{Aker2017StanceCI} 
    &
    \textbf{Structural Features} \cite{Aker2017SimpleOS} \cite{Pamungkas2018StanceCF} \cite{Zubiaga2016StanceCI} 
    \\
    \hline
    \begin{itemize}[leftmargin=0.3cm,topsep=0pt]
        \setlength\itemsep{0.2em}
        \vspace{-0.3em}
        \item \textbf{Word Embedding}: The way of representing words in a tweet such as `Bag of Words' and `Brown Cluster'.
        \item \textbf{POS Tags}: Label a word according with a particular part of speech.
        \item \textbf{Use of Negation}: If the text contains negating words.
        \item \textbf{Swear Words}: If the text contains swear words.
        \item \textbf{Named Entity}: Special names like person, organization, location, etc.
        \item \textbf{Sentiment}: A degree of negativity and positivity can be determined for the tweet. \item \textbf{Surprise Score}: word embedding for surprised words used in the tweet.
        \item \textbf{Doubt Score}: word embedding for doubt words used in the tweet.
        \vspace{-0.7em}
    \end{itemize}
    &
    \begin{itemize}[leftmargin=0.3cm,topsep=0pt]
        \setlength\itemsep{0.2em}
        \vspace{-0.3em}
        \item \textbf{Retweet Count}: Number of retweet for each tweet.
        \item \textbf{URL Count}: Number of URLs in the tweet.
        \item \textbf{Hashtag}: The hashtags used in the tweet.
        \item \textbf{User Mention}: If the tweet mentions users.
        \item \textbf{Question Mark Count}: Number of question marks in the tweet.
        \item \textbf{Exclamation Mark Count}: Number of exclamation marks in the tweet. 
        \item \textbf{Text Length}: Length of the words or characters in a tweet after removing Twitter markers like mentions and hashtags.
        \item \textbf{Favorite Score}: Average number of favorites w.r.t. to the days the user has been active.
        \item \textbf{Capital Ratio}: The ratio of capital letters to all letters in the tweet.
        \item \textbf{Image and Video Count}: Number of images/videos posted by the user.
        \vspace{-0.7em}
    \end{itemize}
    \\
    \hline
    \textbf{Contextual Features} \cite{Zubiaga2018DiscourseawareRS} \cite{Hamidian2019RumorDA}
    &
    \textbf{Conversation Based Features} \cite{Pamungkas2018StanceCF} 
    \\
    \hline
    \begin{itemize}[leftmargin=0.3cm,topsep=0pt]
        \setlength\itemsep{0.2em}
        \vspace{-0.3em}
        \item \textbf{Is Reply}: If the tweet is a reply to its previous tweet.
        \item \textbf{Is Source Tweet}: If the tweet is the source tweet.
        \item \textbf{Is Source User}: If the user is the user who has started the thread.
        \item \textbf{Time Difference}: Difference time between tweets.
        \item \textbf{Event}: Extract the event that the rumor is discussing.
        \vspace{-0.7em}
    \end{itemize}
    &
    \begin{itemize}[leftmargin=0.3cm,topsep=0pt]
        \setlength\itemsep{0.2em}
        \vspace{-0.3em}
        \item \textbf{Tweet Depth}: The number of nodes from the source tweet to the current tweet.
        \item \textbf{Text Similarity to Source Tweet}: Similarity w.r.t. source tweet.
        \item \textbf{Text Similarity to Replied Tweet}:  Similarity w.r.t. previous tweet in the thread.
        \vspace{-0.7em}
    \end{itemize}
    \\
    \hline
    \textbf{User Profile} \cite{Liu2015RealtimeRD} \cite{Zeng2020DetectingRO} \cite{Varshney2021ARO} 
    &
    \textbf{Affective Based Features} \cite{Pamungkas2018StanceCF} \cite{finetunelongformer}
    \\
    \hline
    \begin{itemize}[leftmargin=0.3cm,topsep=0pt]
        \setlength\itemsep{0.2em}
        \vspace{-0.3em}
        \item \textbf{Followers}: Number of followers.
        \item \textbf{Followees}: Number of followees.
        \item \textbf{Friends}: Number of friends.
        \item \textbf{Gender}: If the user is male or female.
        \item \textbf{Historical Microblogs}: Number of tweets posted.
        \item \textbf{Registration Time}: The time that account is created.
        \item \textbf{Location}: If the profile has a location.
        \item \textbf{Profile Has Person Name}: Whether the name of the user profile is a person name.
        \item \textbf{Verification}: If the user is verified.
        \item \textbf{Average Tweet}: Average number of tweets per day.
        \item \textbf{Repost Count}: Number of reposted tweets.
        \vspace{-0.7em}
    \end{itemize}
    &
    \begin{itemize}[leftmargin=0.3cm,topsep=0pt]
        \setlength\itemsep{0.2em}
        \vspace{-0.3em}
        \item \textbf{EmoLex}: Identifying eight primary emotions.
        \item \textbf{Linguistic Inquiry and Word Count (LIWC)}: A psycho-linguistic resource comprising emotional categories.
        \item \textbf{Dictionary of Affect in Language}: Scores for pleasantness, activation, and imagery.
        \item \textbf{Affective Norms for English Words}: English words are rated based on Valence-Arousal-Dominance model.
        \item \textbf{Dialogue-Act Features}: Indicative of communicative goals such as agree, reject, and info-request.
        \item \textbf{Speech-Act Features}: Containing verbs like demand, ask, and report that help categorize stances.
        \vspace{-0.7em}
    \end{itemize}
    \\
    \hline
    \end{tabular}
    \label{tab:featureset}
    \vspace{-0.2em}
\end{table*}

\subsection{Real-Time System}
\fontdimen2\font=0.52ex
Real-time systems are created for the users to search rumorous claims and be presented with the analysis of an automatic rumor detection system. Real-time systems collect and process data in real-time and do not necessarily need early detection, but early detection techniques can be applied in order to make them more efficient.

Liu \textit{et al.} \cite{Liu2015RealtimeRD} create a dynamic system for the Twitter platform that collects event-based tweets (rumorous story and relevant tweets) and extracts event-based practical features in real-time to predict the veracity of rumors. Before manual verification by journalists, their algorithm detects rumors with an accuracy of $85\%$. According to Xu \textit{et al.} \cite{10.1007/978-3-030-03520-4_3}, proposed rumor detection systems are ineffective, and regardless of the outcomes of these systems, users will search websites and news items to verify the reality of rumors on their own. They develop a website where users can enter a Twitter hashtag and receive classified stances of tweets and news items connected to that claim that the system has previously derived. Similarly, Dharod \textit{et al.} \cite{dharod2021trumer} suggest a system that needs users to submit a tweet URL, after which the system extracts local features from the tweet and identifies stances of related tweets, after which the features are fed into a neural network to forecast the veracity of the rumor.

\section{Feature Sets}
\label{sec:featuresets}
\fontdimen2\font=0.52ex
In the classification of stances, features are quite important. More informative features enhance model performance greatly, according to studies \cite{Pamungkas2018StanceCF} \cite{Aker2017SimpleOS}. This section is devoted to the examination of various types of features. The categories and their characteristics are described below, and Table~\ref{tab:featureset} summarizes a large number of instances for each category.

We provide six categories for features based on rumor stance classification studies: \textit{textual}, \textit{structural}, \textit{contextual}, \textit{conversation-based}, \textit{user profile}, and \textit{affective-based} features. In multiple researches, a variety of categories for features have been described. These categories may overlap, implying that the same features may be represented in different categories depending on the perception of the individual. As a result, we present our own categorization of features based on existing research. Table~\ref{tab:featureset} highlights the most commonly used features in research, however it does not include all of the categories of features proposed thus far. As indicated in Section~\ref{sec:conventional}, a rich feature set can increase the performance of a classifier greatly, and future researchers are encouraged to explore into rumor, user, and social media characteristics in order to obtain empirical and valuable features. The six feature categories mentioned above are described below:\vspace{-0.2em}
\begin{itemize}
    \item \textit{Textual features}: Lexicon information, such as the use of negation words, or part of speech (POS) tags, are derived from the text in the tweets. Local features, which are the most commonly used features in these studies, include textual features. Since they include the least amount of information that can be extracted from a tweet, the authors of \cite{Zubiaga2018DiscourseawareRS} use local features to fairly compare the performance of their proposed sequential classifiers with non-sequential classifiers. The model is then enhanced using advanced features to increase the classifier's performance.
    \item \textit{Structural features}: Some Twitter-specific characteristics, such as the amount of hashtags, emoticons, and URLs, are taken into consideration by these features \cite{Pamungkas2018StanceCF, Aker2017SimpleOS}. Structural features can reveal important information about the user's stance as well as whether the rumor is real or not. The amount of question marks, for example, is a key feature of the `query' stance. If a credible individual retweets the rumor, the retweet count could be quite useful in verifying it.
    \item \textit{Contextual features}: These features look into the context of the tweet, or the relationship between the tweet and the tweets around it. Such features include the similarity of the contents of distinct tweets and the stance of the tweet in a thread (source or reply tweets) \cite{Zubiaga2018DiscourseawareRS}.
    \item \textit{Conversation-based features}: These features are suggestive characteristics of a conversational tree and are taken from the tree-structure conversation thread. They can indicate the depth of the tweet in a sequence of tweets and the similarities between tweets. Although similarity between tweets can be regarded a contextual feature, it is also considered a conversation-based feature in some studies such as \cite{Pamungkas2018StanceCF}. We classify the similarity between tweets as a conversation based feature.
    \item \textit{User profile}: Number of followers, gender, registration time, and the number of tweets posted are all included in the user profile \cite{Zeng2020DetectingRO}. User profile information can help in determining the importance score that should be assigned to a user. For example, a user with a large number of followers can have a significant impact on the spread of a rumor.
    \item \textit{Affective-based features}: These features are part of sentiment analysis, which can detect affect-related events \cite{NovAffectDA}. In rumor stance classification, affective elements inspired by their significance in dialogue-act recognition are being used \cite{Pamungkas2018StanceCF}.
\end{itemize}

\section{Datasets}
\label{sec:datasets}
\fontdimen2\font=0.52ex
After annotated datasets were made publicly available, the study of rumor detection flourished over the years. There are datasets specific to some studies that were collected with relation to their specific research topic, in addition to public datasets \cite{Liu2015RealtimeRD, 8923657, Lillie2019JointRS}. Figure~\ref{fig:paperproportation} shows an increase in the number of publications published between 2017 and 2019. This could be owing to the inclusion of a rumorEval shared task with a dataset for stance classification (Task A) and a dataset for veracity detection in the SemEval competitions (Task B). In this section, we first give a summary of both public and private datasets before exploring the limits of rumor-stance databases.
\begin{figure}[!t]
    \begin{tikzpicture}
        \pie[
            color = {
                    yellow!90!black, 
                    green!60!black, 
                    blue!60, 
                    red!70,
                    gray!70,
                    teal!20},
                    text = legend
        ]
        {5/2016, 12/2017, 8/2018, 23/2019, 25/2020, 27/2021}
    \end{tikzpicture}
    \caption{Proportion of papers on rumor stance from years $2016$ to $2021$.}
    \label{fig:paperproportation}
\end{figure}
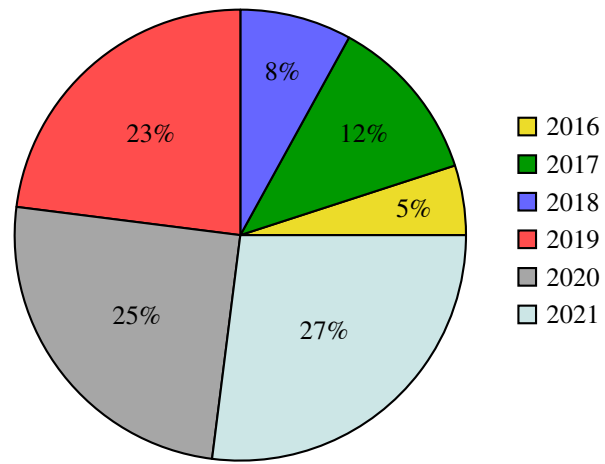

PHEME was a project that began in 2014 with the goal of leveraging veracity detection on social media \cite{derczynski2014pheme}. Zubiaga \textit{et al.} assemble and annotate the PHEME dataset for rumor detection in order to speed up the study of rumor dynamics and propagation \cite{Zubiaga2016AnalysingHP}. \textit{Ferguson unrest}, \textit{Ottawa shooting}, \textit{Sydney siege}, \textit{Charlie Hebdo shooting}, \textit{Germanwings plane crash}, \textit{Prince to play in Toronto}, \textit{Gurlitt collection}, \textit{Putin missing}, and \textit{Michael Essien contracted Ebola} are among the nine newsworthy events gathered from Twitter in the PHEME dataset. It contains a corpus of $4,842$ tweets about the above-mentioned events, including $4,560$ tweets in English and $282$ tweets in German, totaling $330$ rumorous dialogues in a tree-structured format. Based on the evidence in the response tweets, the rumored discussions are annotated. While journalists were able to identify a resolving tweet for some of the rumors, there were none for others in the time frame of the collected dataset.

Due to the challenges of veracity resolution, SemEval-2017 Task 8 was designed. The data is divided into training and test data and includes a part of the annotated PHEME dataset as well as two new events for test data. The training data contains $297$ conversation threads related to eight events, whereas the test data has $28$ threads that have been annotated further. Data for two new events, \textit{Hillary Clinton} and \textit{Marina Joyce}, were collected as part of the test data; the two events make up $8$ of the $28$ threads in the test data; the remaining $20$ threads are related to the same events in the training data. They also included context data, which included information from two sources: (1) Wikipedia articles and (2) the content of linked URLs; they also provided a version of the context that was available prior to the posting of the source tweet. However, only one competitor in the competition used context data \cite{semeval2019}.

The increasing research interest in rumor detection, the inadequacy of datasets, and the significant necessity to debunk misleading claims for it imposes potential harm on society drove SemEval-2019 Task 7 \cite{semeval2019}. rumorEval-2019 has a larger dataset than rumorEval-2017 due to the addition of data from Reddit and new Twitter postings, as well as the promotion of applying subtask A output to subtask B in rumorEval-2019. As training data, the rumorEval-2017 dataset was used, along with additional data from Twitter and Reddit. rumors about natural disasters were collected as test data since rumors can add to the widespread turmoil during natural catastrophes. Task A has a total of $8,574$ data points, of which $6,702$ are training data and $1,872$ are test data, and task B has $446$ data points, of which $365$ are training data and $81$ are test data.

Emergent is a dataset derived from a digital journalism project for rumor debunking \cite{FerrEmergent}. They gathered 300 rumor claims from various sources, including websites, and journalists then gathered related pieces and summarized them into a headline. As a result, the aim is to classify the stance of news headlines and predict the veracity of rumors based on that classification. The category for stances are \textit{for}, \textit{against}, and \textit{observing}, with \textit{for} serving as \textit{support}, \textit{against} acting as \textit{deny}, and \textit{observing} serving as \textit{comment}.

There are multiple datasets acquired by individual researchers in various languages from platforms such as Twitter, Sina Weibo, and Reddit in addition to the aforementioned public databases. Zeng \textit{et al.} \cite{8923657} compiled a user stance Chinese dataset, which is publicly available on github. More than $200,000$ microblogs from Sina Weibo, a Chinese microblogging platform, are included in this dataset. The users' stances are divided into three categories: \textit{support}, \textit{deny}, and \textit{query}. The \textit{comment} label was left out because it resembles tacit support. Lillie \textit{et al.} \cite{Lillie2019JointRS} used Reddit's Danish subreddit to generate a Danish rumor and stance dataset. The data is publicly available and includes $33$ events with a total of $3,007$ Reddit posts. Evrard \textit{et al.} \cite{Evrard2020FrenchTC} have compiled a French corpus that is used for various stance detection tasks. Two annotation systems have been used: (1) the tweet's stance in relation to its prior tweet, and (2) the tweet's stance in relation to the source tweet (rumorous post). To fully cover all conceivable illustrations of tweets' stances, they added a class entitled \textit{ignore} to the standard four classes of rumor stance classification assignment. Albeit they report that the extra label appears to increase the disagreement between annotators. The paper discusses the relationship to the corpus as well as the tools used to generate and annotate the dataset. Another dataset for rumor stance classification introduced by Lozhnikov \textit{et al.} \cite{10.1007/978-3-030-14687-0_16} comprises Russian texts from three sources, namely Twitter, Meduza, and Russia Today. They go on to say that Meduza's remarks are sarcastic, less aggressive, and more grammatically correct than tweets, and that words from Russia Today words are noisy.

Natural language processing on Iberian languages, such as Spanish, Portuguese, Catalan, Basque, and Galician, is the subject of IberEval workshops \cite{ibereval_2017_2022}. IberEval makes a variety of datasets available, including the Catalan dataset for target-specific stance detection. There however, is no rumors stance classification task in IberEval that we are aware of. Furthermore, rumors were obtained using websites such as \texttt{snopes.com}, \texttt{emergent.info}, and \texttt{politifact.com}, and then relevant microblogs, such as users' comments, were gathered using social media platforms such as Twitter API \cite{Liu2015RealtimeRD, semeval2019}. We discovered that several studies have modified or developed already available real-world datasets to match the needs of their research \cite{inproceedings-twitter15&16}.

\begin{table}[!t]
    \centering
    \caption{Distribution of English datasets}
    \setlength{\tabcolsep}{5pt}
    \setlength{\tabcolsep}{3pt}
    \renewcommand{\arraystretch}{1.3}
    \begin{tabular}{|p{75pt}|p{29pt}|p{28pt}|p{28pt}|p{35pt}|}
        \hline
        \textbf{Reference} & \textbf{Support} & \textbf{Deny} & \textbf{Query} & \textbf{Comment}\\
        \hline\hline
        PHEME \cite{Zubiaga2016AnalysingHP} & 910 & 344 & 358 & 2,907 \\
        \hline
        \textbf{Total}$^{\mathrm{*}}$ = 4,519 & 20\%$^{\mathrm{**}}$ & 8\% & 8\% & 64\% \\
        \hline
        RumorEval-2017 \cite{semeval2017} & 1,004 & 415 & 464 & 3,685 \\
        \hline
        \textbf{Total} = 5,568 & 18\% & 7.5\% & 8.5\% & 66\% \\
        \hline
        RumorEval-2019 \cite{semeval2019} & 1,184 & 606 & 608 & 6,176 \\
        \hline
        \textbf{Total} = 8,574 & 14\% & 7\% & 7\% & 72\% \\
        \hline
        Emergent \cite{FerrEmergent} & 1,238 & 394 & - & 963 \\
        \hline
        \textbf{Total} = 2,595 & 48\% & 15\% & - & 37\% \\
        \hline
        \multicolumn{5}{p{200pt}}{$^{\mathrm{*}}$The volume of the dataset given in the upper row.}\\
        \multicolumn{5}{p{200pt}}{$^{\mathrm{**}}$Approximate proportion of the class in the dataset.}\\
    \end{tabular}
    \label{tab:engdataset}
\end{table}
\begin{table}[t]
    \centering
    \caption{Distribution of non-English datasets}
    \setlength{\tabcolsep}{5pt}
    \setlength{\tabcolsep}{3pt}
    \renewcommand{\arraystretch}{1.3}
    \begin{tabular}{|p{75pt}|p{29pt}|p{28pt}|p{28pt}|p{35pt}|}
    \hline
    \textbf{Reference} & \textbf{Support} & \textbf{Deny} & \textbf{Query} & \textbf{Comment} \\
    \hline\hline
    Chinese Dataset \cite{8923657} & 189,390 & 7,701 & 8,364 & - \\
    \hline
    \textbf{Total} = 205,455 & 92\% & 4\% & 4\% & - \\
    \hline
    German Dataset \cite{Zubiaga2016AnalysingHP} & 48 & 13 & 18 & 203 \\
    \hline
    \textbf{Total} = 282 & 17\% & 5\% & 6\% & 72\% \\
    \hline
    Danish Dataset \cite{Lillie2019JointRS} & 273 & 300 & 81 & 2,353 \\
    \hline
    \textbf{Total} = 3,007 & 9\% & 10\% & 3\% & 78\% \\
    \hline
    Russian Dataset \cite{10.1007/978-3-030-14687-0_16} & 58 & 46 & 192 & 662 \\
    \hline
    \textbf{Total} = 958 & 6\% & 5\% & 20\% & 69\% \\
    \hline
    \end{tabular}
    \label{tab:nonengdataset}
\end{table}
The data is distributed unevenly among the four classes, as illustrated in Tables~\ref{tab:engdataset} and \ref{tab:nonengdataset}. For instance, comments account for $66\%$ and $72\%$ of the rumorEval-2017 and rumorEval-2019 datasets, respectively. Two informative classes namely, \textit{deny} and \textit{query} contain the least amount of data, but \textit{comment} comprises the majority of the datasets while contributing the least to the detection of rumors \cite{Scarton2020MeasuringWC}. In most cases, \textit{support} also has a greater volume of data than the two minority classes, which could be an expected outcome given that individuals are generally more likely to support rumors regardless of their veracity, according to \cite{Zubiaga2016AnalysingHP}. Furthermore, some learning models were unable to predict practically any instance of the \textit{deny} class \cite{Kochkina2017TuringAS, Lozano2017MamaEA, Hamidian2019GWUNA}. Different methodologies have been proposed to improve model performance on minority classes. Pamungkas \textit{et al.} train their model on a balanced dataset in which each class contains the same number of data equal to the smallest class, i.e. \textit{deny} \cite{Pamungkas2018StanceCF}. While accuracy declines when compared to training on the entire dataset, model performance on minority classes improves, highlighting the downside of class imbalance. However, balancing the dataset would present a difficulty for the sequential classifiers because it would break the link between the tweets \cite{Zubiaga2018DiscourseawareRS}. Li \textit{et al.} \cite{Li2020RevisitingRS} employ six re-sampling strategies to a BERT-based classifier and feature-based classifiers in order to improve model performance on informative classes. These methods include SMOTE and random under-sampling. In terms of macro-F1 score and GMR, their strategy can greatly increase performance on minority classes while outperforming most models in the SemEval-2017 and SemEval-2019 tasks. 

Evrard \textit{et al.} \cite{Evrard2020FrenchTC} discuss the issues that underpin the proposed annotation schemes. They emphasize that the four classes \textit{support}, \textit{query}, \textit{deny}, and \textit{comment} are insufficient for the task of stance classification because certain tweets are unrelated to the source tweet or cannot be understood by the annotators. These tweets are commonly regarded as comments, however this is an issue in itself. Another issue is that some tweets contain multiple labels at the same time; for example, a tweet that supports a rumor may also raise a doubt about a part of the narrative; thus, putting these tweets into one class may confuse the classifier. The study conducted in \cite{kumar2021rumor} identified inconsistency in labeling of the dataset. They identify comparable responses that are classified into different classes, although the volume of data with this type of inaccuracy is rather small, accounting for $5.75\%$ of the entire training set.

\section{Evaluation}
\label{sec:evaluation}
\fontdimen2\font=0.52ex
To assess the performance of the learning models, various metrics and approaches have been used. It is critical to select an assessment metric that clearly indicates the performance of the model on all classes in order to appropriately select the best performing system, especially for rumor stance classification because the datasets are very skewed. In this section, we examine the evaluation metrics chosen for the task of rumor stance classification and present the results of some methods described in Sections~\ref{sec:rumorMethod} and \ref{sec:timeconsider} using accuracy and macro F1-score as evaluation metrics.

Initially, accuracy was employed as an evaluation metric for rumor stance classification, however due to data imbalance, it is no longer regarded an acceptable choice for comparing learning models. A dummy classifier, for example, differs slightly from well-performing classifiers in terms of micro-average F1 score (accuracy), indicating the inadequacy of accuracy as a criterion for comparison \cite{Zubiaga2016StanceCI}. In SemEval-2019, the macro-average F1 score was chosen as the evaluation metric to have a deeper understanding of the performances \cite{semeval2019}. Scarton \textit{et al.} \cite{Scarton2020MeasuringWC} describe various evaluation metrics that can better reflect the success of a model on all classes, particularly minority classes. They re-evaluate the models' performances in the rumorEval-2017 and rumorEval-2019 competitions using the newly proposed evaluation criteria, and the models that can predict instances better across all classes, particularly informative classes, receive higher scores. They begin by selecting a set of assessment metrics to compare performance, such as accuracy, macro-F$\beta$, geometric mean, area under the ROC curve, and weighted macro-F$\beta$. The best assessment metrics are then chosen based on a comparison of multiple models, namely GMR and weighted F2, which can better determine the most reliable systems (systems that do not misclassify a substantial portion of \textit{deny} instances). The authors add that assigning weights for classes only based on data distribution is inaccurate for the rumor stance classification problem. This is demonstrated by using empirically calculated weights to better represent that the \textit{deny} and \textit{support} classes are the most significant information. In a separate study, Li \textit{et al.} \cite{Li2020RevisitingRS} compare their models using the macro-F1 score and GMR. They demonstrate that the model with the highest GMR (but not the highest macro-F1 score) outperforms all other models, implying that GMR is a better choice for multi-class problems.
\begin{table*}[t]
    \centering
    \caption{Main Results from Papers Highlighted in Sections~\ref{sec:rumorMethod}}
    \setlength{\tabcolsep}{5pt}
    \setlength{\tabcolsep}{3pt}
    \renewcommand{\arraystretch}{1.3}
    \begin{tabular}{|p{85pt}|p{190pt}|p{60pt}|p{40pt}|p{60pt}|}
    \hline
    \textbf{Reference} & \textbf{Model} & \textbf{Dataset} & \textbf{Accuracy} & \textbf{Macro F1-Score} \\
    \hline\hline
    Aker \textit{et al.} \cite{Aker2017SimpleOS} & Random Forest & PHEME & $79.02\%$ & - \\
    \hline
    Zubiaga \textit{et al.} \cite{Zubiaga2016StanceCI} & Linear CRF & PHEME & $64.6\%$ & $43.3\%$ \\
    \hline
    Zubiaga \textit{et al.} \cite{Zubiaga2016StanceCI} & Tree CRF & PHEME & $65.5\%$ & $44\%$ \\
    \hline
    Bai \textit{et al.} \cite{Bai2020ASA} & Stochastic Attention with CNN & PHEME & $78.68\%$ & $44.47\%$ \\
    \hline
    Zhang \textit{et al.} \cite{9376933} & Multi-Modal Multi-Task Learning & PHEME & - & $41.88\%$ \\ 
    \hline
    Ma \textit{et al.} \cite{10.1145/3184558.3188729} & Enhanced Shared-Layer Multi-Task Learning & PHEME & $62.2\%$ & $43\%$ \\
    \hline
    Pamungkas \textit{et al.} \cite{Pamungkas2018StanceCF} & SVM & SemEval-2017 & $79.5\%$ & $47\%$ \\
    \hline
    Poddar \textit{et al.} \cite{Poddar2018PredictingSI} & Neural Network & SemEval-2017 & $79.86\%$ & - \\
    \hline
    Veyseh \textit{et al.} \cite{10.1145/3132847.3133116} & CNN & SemEval-2017 & $82\%$ & $48.2\%$ \\
    \hline
    Kumar \textit{et al.} \cite{kumar2021rumor} & Capsule Network and Multi-Layer Perceptron & SemEval-2017 & $77.6\%$ & $55\%$ \\
    \hline
    Yu \textit{et al.} \cite{Yu2020CoupledHT} & Hierarchical Transformer with BERT & SemEval-2017 & $76.3\%$ & $50.9\%$ \\
    \hline
    Khandelwal \textit{et al.} \cite{finetunelongformer} & Longformer Transformer with Attention Mechanism & SemEval-2019 & - & $67.2\%$ \\
    \hline
    Scarton and Li \cite{Scarton2021CrosslingualRS} & BERT and Machine Translation (full multilingual model) & Danish dataset & - & $32.3\%$ \\
    \hline
    Scarton and Li \cite{Scarton2021CrosslingualRS} & BERT and Machine Translation (full multilingual model) & Russian dataset & - & $52.4\%$ \\
    \hline
    Giasemidis \textit{et al.} \cite{semisupervisedGias} & Label Spread (semi-supervised algorithm) & Private Dataset & - & $47.63\%$ \\
    \hline
    \end{tabular}
    \label{tabeval1}
\end{table*}

Henceforth, we explore the superiority of recently suggested models over past efforts and highlight specifics of the high-performing models. In \cite{Li2020RevisitingRS}, the authors train four algorithms, namely, random forest, multi-layer Perceptron, logistic regression with stochastic gradient descent, and pre-trained BERT model, with four resampling mechanisms and threshold-moving, namely, random under-sampling and oversampling, SMOTE, ADASYN, and SMOTEENN, using the SemEval-2017 dataset. They compare them using the macro F1-score and GMR to find the model that performs well on the two most informative classes, \textit{deny} and \textit{support}. BERT with no resampling has the highest macro F1-score of $58.4\%$, while BERT with threshold-moving has a GMR of $62.6\%$ and a lower macro F1-score ($54\%$ percent). A greater GMR indicates better performance of \textit{deny} and \textit{support} instances, implying that BERT with threshold-moving is better. Other systems, on average, perform better with random under-sampling. There are other systems with a GMR of $0$, indicating that they are unable to predict any instances of the \textit{deny} class. In terms of macro F1-score, the hierarchical transformer proposed in \cite{Yu2020CoupledHT} using the SemEval-2017 dataset outperforms four previous works, namely, SVM \cite{Pamungkas2018StanceCF}, BranchLSTM \cite{kochkina2018all}, temporal attentional model \cite{10.1145/3132847.3133116}, and conversational graph convolutional Network ($50.9\%$) \cite{Wei2019ModelingCS}. Their classifier can predict \textit{deny} instances substantially better than the baselines, with two of them failing to detect any instances of the \textit{deny} class but achieving competitive results for the \textit{support} class \cite{Pamungkas2018StanceCF, kochkina2018all}. Their multi-task learning model (coupled hierarchical transformer) also surpasses the multi-task learning frameworks in \cite{kochkina2018all, Wei2019ModelingCS}, reaching a macro F1-score of $68\%$ on the SemEval-2017 dataset and $39.6\%$ on the PHEME dataset.

On the SemEval-2016 stance detection dataset, the authors of \cite{Tian2020EarlyDO} train the BERT model and CNN+LSTM with stance transfer learning and compare them in terms of macro F1-score for the task of early rumor detection. The stance transferring methods outperform the non-stance methods, indicating that stances can have a considerable impact on the rumor verification task in terms of early detection of rumors. They also show the efficacy of the BERT model with stance transferring over the BERT model by presenting the clusters of each model for different rumors, namely non-rumor, true rumor, false rumor, and unverified rumor, illustrating that the BERT model with stance transferring can very well divide rumors into distinguishable clusters. To determine the superiority of their proposed deep learning method, Kumar \textit{et al.} \cite{kumar2021rumor} compare their combined capsule and multi-layer Perceptron neural networks to CNN-based baselines. CNN with multi-layer Perceptron exceeds simple CNN and capsule neural network, which have macro F1-scores of $26\%$ and $27\%$, respectively, while a combination of CNN and capsule neural network outperforms all other CNN-based baselines with a $55\%$ macro F1-score. Multi-task multi-modal learning proposed by Zhang \textit{et al.} \cite{9376933} beats prior efforts such as LSTM multi-task learning and GRU multi-task learning for rumor veracity prediction with a macro F1-score of $80.41\%$ and $82.02\%$ on the SemEval-2017 and PHEME datasets, respectively. They demonstrate the effectiveness of the multi-modal post embedding layer by comparing their multi-task framework to non-multi-task models, revealing that their model outperforms the baseline models.

The cross-lingual settings set forth by \cite{Scarton2021CrosslingualRS} show varying results for different languages. For example, when Russian is machine translated into English, the macro F1-score is $46.7\%$, and it is $44.2\%$ otherwise. This is explained by addressing the size of the Russian dataset for training, whereas English data points are also used for training when it is translated into English. Finally, the best-performing models attain a macro F1-score of $35.2\%$ for Danish models, $50.6\%$ for Russian models in few-shot setting, and $46.4\%$ for German models in zero-shot setting. The whole multilingual model yields considerable results for the English language in terms of GMR and wF2, but a reduction in macro F1-score ($48.4\%$). This experiment proves that the multi-lingual BERT Model and machine translation can help languages with fewer datasets and even improve performance in languages with a large number of datasets. For the rumor verification, Khoo \textit{et al.} \cite{khoo2020interpretable} test their post-level attention model on three datasets, namely PHEME and Twitter15\&16. They report a low performance of $39.5\%$ macro F1-score on the PHEME dataset, and explain that this could be related to the event-based train-test split technique. To solve this issue, they randomly separated PHEME data points and trained the model on the new dataset, achieving a $77.4\%$ macro F1-score. Table~\ref{tabeval1} highlights the results of the aforementioned works on rumor stance classification task. This table is not to compare these methods with each other, since the datasets are different, and for those with the same dataset, the splitting methods may differ which could make it unfair for comparison.

\section{Findings}
\label{sec:findings}
\fontdimen2\font=0.52ex
Aside from the ultimate outcomes of its proposed system, each study report offers instructive background that plays a significant role in system design. In our case, we may gain similar findings by studying rumors and social media, as well as analyzing the performance of learning models with various settings, which can give us with insights into the specifics of our method. In this section, we look at the findings of research on rumor. This comprises investigating a dataset or conducting analysis after training the model on a dataset. In addition, we look into efforts that report feature and error analyses in order to better understand the performance of their models.

\subsection{Study of rumor}
\fontdimen2\font=0.52ex
Studying the spread of rumors on social media, the temporal characteristics of responses to a rumor post, and the lifespan of a rumor would provide us with useful information for developing the correct perspective on rumor detection. According to \cite{Poddar2018PredictingSI}, feeding temporal information to a model can help improve performance. Analysis for early rumor detection revealed that $50\%$ of comments appear $60$ minutes after a rumor is posted and $80\%$ appear $100$ minutes later, indicating that comments can be employed for early rumor identification. For example, the authors of \cite{Tian2020EarlyDO} have established $60$ minutes as the default option for early rumor detection and emphasize that user comments contain early signals that can aid in early rumor detection.

According to research, true rumors are validated faster than false rumors, and unverified rumors are widely propagated \cite{Zubiaga2016AnalysingHP}. Users generally support unconfirmed rumors, and news organizations generally support rumors and bring evidence regardless of the reality of the rumors, which may be related to the pressure to produce content and the competitive atmosphere among news organizations \cite{Zubiaga2016AnalysingHP}. Temporal analysis of user activity has yielded intriguing results that can be used to improve rumor identification. For example, users tend to question the rumorous story early in its dissemination, and the number of query stances decreases over time. If, on the other side, the rumor is false, the number of denial stances rises \cite{10.1145/3132847.3133116}. Furthermore, comments make up the majority of tweets that are posted later in time, and the aggregate quantity of \textit{support}, \textit{deny}, and \textit{query} tweets decreases over time, implying that the credibility of the rumor is discussed at the time the rumor is posted \cite{Zubiaga2018DiscourseawareRS}.

The difficulties of verifying a rumor with a small dataset is well shown in \cite{kochkina2018all}. The authors train their model on five big noteworthy events from the PHEME dataset, as well as the complete dataset (9 events). The remaining four events are on a smaller scale and are false or unproven rumors, resulting in a performance drop in the model. They also prove, using Kurtosis and entropy values, that smaller events tend to increase model complexity. In another study, extracting opposing opinions and building a credibility network through supporting and opposing relationships demonstrated relative effectiveness in performance \cite{Jin2016NewsVB}.

\subsection{Feature Analysis}
\fontdimen2\font=0.52ex
Feature analysis provides a deeper understanding of how different features contribute to each user's stance. Each stance class has distinct characteristics that set it apart from the others. For example, the authors of \cite{Zubiaga2018DiscourseawareRS} discovered that \textit{query} stances typically contain question marks, \textit{deny} stances employ negation more frequently than other types of stances, and \textit{support} tweets include a link that is likely to bring evidence. They also showed that leveraging the discursive characteristics of Twitter interactions, such as the probability of transitions within tree-structured conversational threads, can result in significant improvements. Their findings show that the LSTM classifier with a more limited set of features performs best, owing to its ability to handle context natively, as well as relying on branches rather than the entire conversational tree, which reduces the amount of data and complexity that must be processed in each sequence. In their other work, the same authors observe that classifiers struggle to categorize the \textit{deny} class, although this does not always hold true for the \textit{query} class, despite the fact that both are minority classes. This could be because some features, such as the question mark, are unique to the query class, yet denials may resemble the \textit{comment} class, and conversely, comments may contain negating words, making them similar to the \textit{deny} class \cite{Zubiaga2016StanceCI}.

\subsection{Error Analysis}
\fontdimen2\font=0.52ex
Pamungkas \textit{et al.} \cite{Pamungkas2018StanceCF} perform error analysis to discuss possible reasons for misclassification cases of their system. For example, they found that some reply tweets with a \textit{deny} stance show hatred towards the author of the source tweet rather than the rumor content; these tweets were frequently misclassified as \textit{comment} because their system groups \textit{deny} tweets by tweets that refute the content of the source tweet. Their model also failed to categorize some tweets from the \textit{deny} class with brief texts, such as `shut up!', correctly arguing that they do not contain enough information. Another interesting finding was that in rare cases where the source tweet denies the rumorous story and reply tweets that agree with the source tweet, implying disproving the rumor content, use language like \textit{support} class, causing the classifier to label them as \textit{support} when they are, in fact, \textit{deny} stances.

\section{Future Research and Challenges}
\label{sec:future}
\fontdimen2\font=0.52ex
Despite the different approaches proposed for rumor stance classification, more work remains to be done in order to achieve reliable results that may effectively assist the rumor verification system in correctly identifying the veracity of rumors. In this section, we explore some of the shortcomings that need to be filled in order to improve rumor detection on social media platforms.

Because the existing datasets for rumor stance classification are largely in English, non-English social media platforms lack a comprehensive and well-developed mechanism for automatically detecting rumors. As stated in Section~\ref{sec:datasets}, several non-English datasets have been collected for rumor stance classification. Future researchers can utilize these datasets to propose new methods or improve on prior work to create rumor detection systems for additional languages. When a rumor is published online, it can spread across numerous platforms and languages, or the responses can be in a range of languages. Hence, having a multi-lingual model would improve rumor detection. Different modalities, including as text, image, video, and audio, are employed on social media platforms and can be leveraged to acquire more precise results for user stances. Furthermore, datasets on modalities other than text are limited, leaving much work to be done in this field. Another issue with the datasets is that the vast majority of them are obtained from Twitter. The features of different platforms differ, and dealing with data from other platforms may provide new insights into the rumor detection problem.

When a model trained on certain rumors is applied to rumors from another domain, such as health and non-health related rumors, its performance suffers. As a result, feature development can be regarded as a solid research strategy for gaining insights into rumor stance features that are accurate representations of rumors across many domains. Cross-domain approaches must also receive more attention in order to address the problem of performance degradation across domains. Moreover, early discovery of rumors is critical to preventing their spread and the potential harm they bring to society and individuals. Given the little information available at the time a rumor is disseminated, the task of early rumor detection becomes rather difficult. Researchers exploring early detection of rumors have uncovered interesting information on early signals in rumors, such as the fact that the propagation structure of rumors is more complex than that of non-rumors in the time close to the posting of the rumor. More research may be done to determine the best learning model for early rumor detection, and an evaluation of several feature sets can assist better understand the nature of early information in rumors. 

Another issue is that unsupervised and semi-supervised algorithms are less well-known in this field. These are useful methods since labeling data is a time-consuming and labor-intensive task emphasizing the importance of developing non-supervised algorithms. In addition, transfer learning is another approach that overcomes the intense human labor required for data annotation, but little research has been done on this approach.

All research on rumor stance classification is done to improve the accuracy of rumor verification systems, which are then used in real-world scenarios to stop the spread of rumors. As a result, the proposed models and findings must be tested in a real-world system. A rising argument in a real-world system is to inform users on why a rumor recognized as false is false, with the goal of educating users on how to judge the veracity of a rumor. This is a public education programme that teaches people how to identify between false and accurate rumors. In addition to a machine learning algorithm that learns the features of a rumor, humans may also learn how to evaluate a rumorous story. Furthermore, such a framework must dynamically work on rumorous stories and their context, and additional study might be undertaken to improve the performance of a real-time system.

\section{Conclusion}
\label{sec:conclusion}
\fontdimen2\font=0.52ex
In this paper, we provided an inclusive representation of automatic rumor stance classification problem to present practitioners and researchers with the details of studies conducted in this domain. We presented different approaches proposed for rumor stance classification categorized under \textit{traditional machine learning}, \textit{deep learning}, and \textit{early detection}. We then introduced different categories of feature sets employed in rumor stance classification models, followed by a thorough description of English and non-English dataset collections along with their characteristics and differences. Furthermore, we explained the assessment procedures adopted in existing efforts and tabulated the performance evaluations on various datasets for comparative analysis. Finally, to assist interested readers in their quest to close the gaps in this field of research, potential future works on user stance classification and automatic rumor detection were suggested.


\bibliographystyle{IEEEtran}
\bibliography{IEEEabrv,myref}

\end{document}